\documentclass[%
 reprint,
onecolumn,
 amsmath,amssymb,
 aps,
]{revtex4-2}

\usepackage{amsmath}
\usepackage{amssymb}
\usepackage{graphicx}
\usepackage{dcolumn}
\usepackage{bm}

\usepackage{xcolor}

\definecolor{brickred}{rgb}{0.8, 0.25, 0.33}
\definecolor{applegreen}{rgb}{0.55, 0.71, 0.0}
\definecolor{britishracinggreen}{rgb}{0.0, 0.26, 0.15}
\definecolor{acquamarina}{rgb}{0.5, 1.0, 0.83}
\definecolor{albicocca}{rgb}{0.98, 0.7, 0.2}

\newcommand{\mattia}[1]{\textcolor{brickred}{\textbf{#1}}}

\begin{document}

\preprint{APS/123-QED}


\title{Reconstructing higher-order interactions in coupled dynamical systems}

\author{Federico Malizia}
\affiliation{Dipartimento di Fisica ed Astronomia, Università di Catania and INFN, Catania, Italy}
\author{Alessandra Corso}%
 \affiliation{Department of Electrical Electronic and Computer Science Engineering, University of Catania, Catania, Italy}%
 
\author{Lucia Valentina Gambuzza}
 \affiliation{Department of Electrical Electronic and Computer Science Engineering, University of Catania, Catania, Italy}%
\author{Giovanni Russo}
\affiliation{Department of Mathematics and Computer Science, University of Catania, 95125 Catania, Italy}%

\author{Vito Latora}
\affiliation{School of Mathematical Sciences, Queen Mary University of London, London, E1 4NS, United Kingdom\\ Dipartimento di Fisica ed Astronomia, Università di Catania and INFN, Catania, Italy\\
Complexity Science Hub, Josefstäadter Strasse 39, A 1080, Vienna, Austria}

\author{Mattia Frasca}
\affiliation{ Department of Electrical Electronic and Computer Science Engineering, University of Catania, Catania, Italy}

\date{\today}

\begin{abstract}
Higher-order interactions play a key role for the stability and function of a complex system. However, how to 
identify them is still an open problem. Here, we propose a method to fully reconstruct the structural connectivity of a system of coupled dynamical units, identifying both pairwise and higher-order interactions from the system time evolution. Our method works for any dynamics, and allows the reconstruction of both hypergraphs and simplicial complexes, 
either undirected or directed, unweighted or weighted. With two concrete applications, we show how the method can help understanding the ecosystemic complexity of bacterial systems, or the microscopic mechanisms of interaction underlying coupled chaotic oscillators. 
\end{abstract}

\maketitle

\section{Introduction}

Higher-order interactions are present in ecosystems, where the way two species interact can be influenced by a third species \cite{grilli2017higher}, in social systems, where interactions in groups of three or more individuals naturally occur \cite{iacopini2019simplicial}, in the brain cortex \cite{yu2011higher}, and in  many other complex systems  \cite{battiston2020networks}. 
Recent studies based on mathematical tools such as simplicial complexes \cite{bianconi2021higher,hatcher2005algebraic} and hypergraphs \cite{berge1973graphs} have already demonstrated that the dynamics in presence of higher-order interactions can be significantly different from that of systems where interactions are exclusively pairwise \cite{iacopini2019simplicial,alvarez2021evolutionary,gambuzza2021stability,gallo2022synchronization,muolo2023turing}. 
%
%
How to infer and model higher-order interactions is then crucial for understanding the dynamics and functioning of complex systems \cite{battiston2021physics,battiston2022higher}. While in complex networks, the reconstruction problem, also known as the inverse problem, i.e. determining the network from the dynamics of a system,  has been dealt with different techniques \cite{timme2014revealing}, the question on how to infer connectivity in presence of higher-order interactions is still open. 

Concerning reconstruction in complex networks, two different types of approaches, which target either the functional or the structural connectivity of the system, have been developed. 
In the first type of approaches, functional networks are typically constructed from the network temporal evolution by evaluating correlation, Granger causality or transfer entropy among the signals of the different network units, or using Bayesian inference methods \cite{ren2010noise,wu2011detecting,jansen2003bayesian}. 
In the second type of approaches, the underlying structural connectivity of a network is obtained from the network response to external perturbations \cite{timme2007revealing}, from its  synchronization with a copy with adaptive links \cite{yu2006estimating,wu2015identifying}, or from the solution of optimization problems based on measurements of node time series, when the functional form of the node dynamics is known \cite{shandilya2011inferring,han2015robust,shi2021inferring}. 

The reconstruction problem in the presence of 
higher-order interactions is more convoluted. Recently, the fundamental distinction between {\em higher-order mechanisms}, i.e.~the presence of higher-order terms in the microscopic structure of the interactions, 
and {\em higher-order behaviors}, i.e.~the emergence of higher-order correlations in the dynamical behaviour of a system, has been pointed out \cite{rosas2022disentangling}.
The relationship between these two is not trivial as higher-order behaviors do not necessarily rely on higher-order mechanisms. However, for their identification, techniques that go beyond pairwise statistics are required. 
%
%
For instance, information-theoretic approaches to study multivariate time series (of node activities) based on hypergraphs \cite{marinazzo2022information}, higher-order predictability measures (such as generalizations of Granger causality and partial information decomposition) \cite{pernice2022pairwise}, or simplicial filtration procedures  \cite{santoro2023unveiling} have been proposed to extract important information on higher-order behaviours that otherwise would not be visible to standard, i.e., network-based, analysis tools.
%
Higher-order behaviors, which are likely due to the presence of higher-order mechanisms, can be identified by recently introduced techniques for statistical validation 
able to detect
overexpressed hyperlinks \cite{musciotto2021detecting,musciotto2022identifying}. Other statistical approaches to the problem are based on Bayesian methods, and have been used to construct hypergraphs directly from pairwise measurements (link activities), 
%
even in cases where the higher-order interactions are not explicitly encoded \cite{young2021hypergraph,lizotte2022hypergraph}. 
%
Statistical inference and expectation maximization are also at the basis of a method recently developed to reconstruct  higher-order mechanisms of interaction in simplicial SIS spreading 
and Ising Hamiltonians with two- and three-spin interactions \cite{wang2022full}. However, this method can only be applied to binary time series data produced by discrete two-state dynamical models.

In this work, we propose an optimization-based approach 
to infer the high-order structural connectivity of a complex system from its time evolution, which works in the case of the 
most general continuous-state dynamics, i.e.~when node variables are not restricted 
to take binary values. 
Namely, we consider a system made by many dynamical units (nodes) coupled through pairwise and higher-order interactions; we assume that the local dynamics and the functional form of interactions are known \cite{prasse2022predicting} or identifiable \cite{keesman2011system,gallo2022lack}, and we propose a method to extract  
the topology of such interactions by solving an optimization problem based on the measurement of the time evolution of the node variables.
With two concrete applications, we show that the method can effectively reconstruct which nodes are interacting in pairs and which in groups of three or more nodes. 

\section{Reconstruction of the interactions}

As a general model of a  dynamical system of $N$ nodes coupled through pairwise and higher-order interactions, we consider the following set of equations:
\begin{equation}
\label{eq:general_dynamics}
\begin{array}{lll}
\dot{\mathbf{x}}_i &=& \mathbf{f}_i(\mathbf{x_i}) + \sum\limits_{d=1}^D \sigma_d\sum\limits_{j_1,\dots,j_d=1}^N a_{ij_1\dots j_d}^{(d)}\mathbf{g}^{(d)}(\mathbf{x}_i,\mathbf{x}_{j_1},\dots,\mathbf{x}_{j_d}), 
\end{array}
\end{equation}
with $i=1,\ldots,N$.
Here $\mathbf{x}_i(t)\in\mathbb{R}^{m}$ is the state vector of unit $i$, $\mathbf{f}_i:\mathbb{R}^{m}\rightarrow \mathbb{R}^m$ is the nonlinear function describing the local dynamics at node $i$, while $\mathbf{g}^{(d)}:\mathbb{R}^{m\times (d+1)}\rightarrow \mathbb{R}^m$ are the nonlinear functions of order $d$, modeling interactions in groups of $d+1$ nodes, with $d=1,\ldots,D$.
The topology of $(d+1)$-body interactions is encoded in the tensor $A^{(d)}$, while the parameters 
$\sigma_1$, $\dots$, $\sigma_D > 0$ tune the strengths of the coupling at each order. 

Here we want to infer the complete 
structural connectivity of a dynamical system, which means we want to reconstruct, not only the entries of the adjacency matrix $A^{(1)} = \{  a^{(1)}_{ij} \}  $ from the knowledge of the evolution of the state variables $\mathbf{x}_1(t),\ldots,\mathbf{x}_N(t)$, 
but also the higher-order interactions encoded by 
the tensors $A^{(2)} = \{  a^{(2)}_{ijk} \}  ,\ldots,A^{(D)}=\{ a^{(d)}_{ij_1 \ldots j_d}\}$. 
In doing this, we 
assume that the functions $\mathbf{f}_i$ and $\mathbf{g}^{(1)}, \mathbf{g}^{(2)} \ldots, \mathbf{g}^{(D)}$, 
are known. This is a reasonable assumption as the local dynamics of many real-world complex systems, as well as the functional forms of their interactions, have been well identified.
For instance, well-established mathematical models to describe the dynamics of neurons and synapses, or the growth of a biological species when in isolation, or when in interactions with other species, are available. 
In absence of such models, we assume instead that, prior to the structural connectivity reconstruction, the model of the isolated dynamics of a single unit, a pair, a group of three units, etc. can be derived using proper identification techniques \cite{keesman2011system}. 

Our reconstruction technique works as follows. 
Suppose we have access to $M$ measurements of 
the variables $\mathbf{x}_1(t),\ldots,\mathbf{x}_N(t)$ at times $t$  respectively equal to
$h \Delta T$ with $h=1,...,M$, 
and $\Delta T$ being a (constant) sampling interval.
The discretized version of Eqs.~\eqref{eq:general_dynamics} reads:
\begin{equation}
\label{eq:general_dynamicsTD}
\begin{array}{l}
\frac{\mathbf{x}_i(h+1)-\mathbf{x}_i(h)}{\Delta T} = \mathbf{f}_i(\mathbf{x_i}(h)) +\sum\limits_{d=1}^D \sigma_d\sum\limits_{j_1,\dots,j_d=1}^N a_{ij_1\dots j_d}^{(d)}\mathbf{g}^{(d)}(\mathbf{x}_i(h),\mathbf{x}_{j_1}(h),\dots,\mathbf{x}_{j_d}(h)),
\end{array}
\end{equation}
\noindent where $i=1,\ldots,N$, $h=1,\ldots,M-1$, and $\mathbf{x}_i(h)$ is a short notation for $\mathbf{x}_i(h \Delta T)$. 

Now, let $\mathbf{y}_i(h):=\frac{\mathbf{x}_i(h+1)-\mathbf{x}_i(h)}{\Delta T} -\mathbf{f}_i(\mathbf{x_i}(h))$ and $\mathbf{Y}_{i}=[ \mathbf{y}_{i}(1)^T,\ldots,\mathbf{y}_{i}(M)^T]^T$.
Let us define the vector 
\begin{equation}
\nonumber
\begin{array}{l}
\mathcal{A}_{i} 
\equiv 
[(\mathcal{A}^{(1)}_i)^T, (\mathcal{A}^{(2)}_i)^T, 
\ldots (\mathcal{A}^{(D)}_i)^T]^T =\\ \quad
=[ a^{(1)}_{i1},\ldots,a^{(1)}_{i,i-1},a^{(1)}_{i,i+1}, \ldots, a^{(1)}_{iN},  a^{(2)}_{i12},\ldots,a^{(2)}_{i,N-1,N}, \ldots,a^{(D)}_{i,1,\ldots,D},\ldots,a^{(D)}_{i,N-D+1,N}]^T.\end{array}
\end{equation}
which contains the quantities we want to reconstruct for each node $i$.
Then, from Eqs.~(\ref{eq:general_dynamicsTD}), we get:
\begin{equation}
\label{eq:linearsys}
\mathbf{Y}_i=\Phi_i \mathcal{A}_{i}
\end{equation}
\noindent for $i=1,\ldots,N$, with
\begin{widetext}
\begin{equation}
\label{eq:Phi}
\Phi_i =  \left [ 
\begin{array}{ccccccccc}
    \sigma_1\mathbf{g}^{(1)}_{i,1}(1) & \ldots & \sigma_1\mathbf{g}^{(1)}_{i,i-1}(1) & \sigma_1\mathbf{g}^{(1)}_{i,i+1}(1) & \ldots & \sigma_1\mathbf{g}^{(1)}_{i,N}(1) & \sigma_2\mathbf{g}^{(2)}_{i,1,2}(1) & \ldots &
    \sigma_D\mathbf{g}^{(D)}_{i,N-D+1,\ldots,N}(1)
    \\
    \vdots & & & & & & & & \vdots\\
    \sigma_1\mathbf{g}^{(1)}_{i,1}(M) & \ldots & \sigma_1\mathbf{g}^{(1)}_{i,i-1}(M) & \sigma_1\mathbf{g}^{(1)}_{i,i+1}(M) & \ldots & \sigma_1\mathbf{g}^{(1)}_{i,N}(M) & \sigma_2\mathbf{g}^{(2)}_{i,1,2}(M) & \ldots &
    \sigma_D\mathbf{g}^{(D)}_{i,N-D+1,\ldots,N}(M)
    \\
\end{array}
\right ]
\end{equation}
\end{widetext}
where we introduced the following short notation: 
$\mathbf{g}^{(d)}_{i,j_1,\ldots,j_d}(h):=\mathbf{g}^{(d)}(\mathbf{x}_i(h),\mathbf{x}_{j_1}(h),\dots,\mathbf{x}_{j_d}(h))$.

For each node $i$ we need to identify $H=N-1+(N-1)(N-2)+\ldots+(N-1)\cdots(N-D)$ terms, corresponding to the entries of $\mathcal{A}_{i}$. Therefore, $\Phi_i \in \mathbb{R}^{M \times H}$. Solving Eq.~(\ref{eq:linearsys}) for the unknowns $\mathcal{A}_{i}$ allows one to reconstruct all interactions of node $i$, such that the whole structural connectivity can be inferred by repeating the calculations for all nodes, $i=1,\ldots,N$. Notice that Eq.~(\ref{eq:linearsys}) maps the problem of the reconstruction of the higher-order interactions into that of solving a system of algebraic equations in the unknown variables given by the $H$ entries of $\mathcal{A}_{i}$.

Notice that, when $M<H$, the system of equations~(\ref{eq:linearsys}) is underdetermined and multiple solutions may exist~\cite{timme2014revealing}. 
Conversely when $M>H$, the system of equations~(\ref{eq:linearsys}) is overdetermined and can be solved using the method of least squares~\cite{timme2014revealing}. 
%
%
%

We will now show that our approach is able to successfully reconstruct the full set of interactions at any order in the case of two completely different dynamics, namely microbial ecosystems and systems of coupled chaotic oscillators.
The considered case studies will also demonstrate that the method works for the reconstruction of both hypergraphs and simplicial complexes, and no matter whether the underlying structure is undirected or   directed, unweighted or weighted. 

\section{Lotka-Volterra dynamics}

\begin{figure*}[t]
\begin{center}
\includegraphics[width=\linewidth]{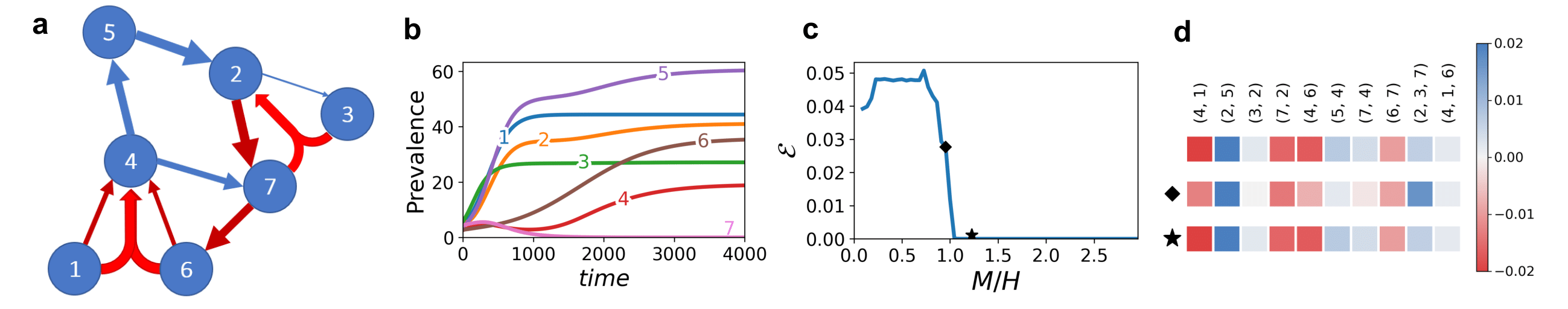}
\caption{Reconstructing higher-order interactions in a microbial ecosystem. (a) The underlying weighted hypergraph of a Lotka-Volterra system with $N=7$ species and 2- and 3-body interactions, which we want to reconstruct from (b) the time evolution of the seven species abundance $x_i(t)$. (c) Quality of the reconstruction is measured by reporting the error $\mathcal{E}$ as a function of the ratio between the length $M$ of the trajectories and the number $H$ of interactions to reconstruct. (d) The reconstructed weights (color coded) for each 2- and 3-body interaction are compared to the true weights, for the two cases 
$M/H = 0.95$ ($\blacklozenge$) and 
$M/H = 1.23$ ($\star$), also indicated in panel (c).
\label{fig:figure1}}
\end{center}
\end{figure*}

In our first application we focus on the dynamics of microbial ecosystems. These consist of species that may engage in diverse relationships, either cooperative, such as the transfer of complementary metabolites, or antagonistic, such as competition for a resource \cite{hibbing2010bacterial}. 
The validation of community-wide interactions in microbial communities is a far from trivial problem, faced both with experimental approaches \cite{trosvik2010web} and through the use of mathematical modeling \cite{berry2014deciphering}. The problem is further complicated by potential higher-order interactions, which play a role in stabilizing diverse ecological communities and maintaining species coexistence \cite{grilli2017higher,aladwani2019addition,singh2021higher}. Here, we model a microbial ecosystem of $N$ species as a hypergraph of $N$ coupled Lotka-Volterra equations \cite{case1999illustrated} including both pairwise and three-body interactions:  
%
\begin{equation}
\label{eq:LotKaVolterra}
\dot{x}_{i} = f_i (x_i)
+ \sum\limits_{j=1}^N a_{ij}^{(1)} x_i x_j +\sum\limits_{j = 1}^N\sum\limits_{k=1}^N a_{ijk}^{(2)} x_i x_jx_k  
\end{equation}
with $i=1,\ldots,N$.  
%
%

The variable $x_i$ represents the abundance of species $i$. The local dynamics of $x_i$ is governed by the logistic function ${f}_i(x_i)=r_i x_i \left ( 1-\frac{1}{k_i}x_i  \right )$ where $r_i$ and $k_i$ are the growth rate and the carrying capacity. The pairwise interactions between species are encoded in the coefficients of the $N\times N$ weighted matrix  $\mathrm{A}^{(1)}=\{a_{ij}^{(1)}\}$, while the three-body interactions in the coefficients of the $N\times N \times N$ weighted tensor $\mathrm{A}^{(2)}=\{a_{ijk}^{(2)}\}$. Note that Eqs.~\eqref{eq:LotKaVolterra} are in the form of Eqs.~\eqref{eq:general_dynamics} with ${g}^{(1)}(x_i,x_j)=x_ix_j$ and ${g}^{(2)}(x_i,x_j,x_k)=x_ix_jx_k$. As an example, we consider the system of $N=7$ species 
with four cooperative 
 ($a_{ij}^{(1)}>0$) and four antagonistic ($a_{ij}^{(1)}<0$) pairwise interactions,
studied in Ref.~\cite{berry2014deciphering}
and shown in Fig.~\ref{fig:figure1}(a) with blue and red arrows respectively. 
In addition to these pairwise interactions, we have included two antagonistic three-species interactions, shown as double arrows in the hypergraph in Fig.~\ref{fig:figure1}(a). These respectively correspond to a contribution to the dynamics of $x_2$ given by $a_{237}^{(2)}x_2x_3x_7$ 
and one to $x_4$ given 
by $a_{416}^{(2)}x_4x_1x_6$, with $a_{237}^{(2)}=-0.0062$ and $a_{416}^{(2)}=-0.0016$
\cite{grilli2017higher}.
The other system parameters have been set as in~\cite{berry2014deciphering}. Namely, growth rates $r_i$ for all species have been selected from a uniform distribution in the interval $(0,1)$, similarly the carrying capacities $k_i$ from a uniform distribution in the interval $(1,100)$, and the initial conditions in the interval $(10,100)$. 

Under these conditions, as shown by the time evolution of the variables $x_i(t)$, with $i=1,\ldots,7$, reported in Fig.~\ref{fig:figure1}(b), the microbial ecosystem typically converges to a stable equilibrium point corresponding to the coexistence of six species over seven. To feed our reconstruction algorithm, we sampled the time evolution of $\{ x_i(t) \}_{i=1,\ldots,7}$ at $M$  regular intervals of size $\Delta T=0.01$, and we then used these values to calculate $\mathbf{Y}_{i}$ and $\mathrm{\Phi}_i$ from Eq.~(\ref{eq:Phi}). At this point, optimization via the method of least squares produces, for each $i=1,\ldots,7$, a vector $\hat{\mathcal{A}}_{i}$ that is solution of Eq.~\eqref{eq:linearsys}. To quantify the accuracy of the reconstruction of the interactions at any order, we compare the estimation $\hat{\mathcal{A}}_{i}$ with the true values of the couplings, $\mathcal{A}_{i}$, for each $i$, evaluating the reconstruction error $\mathcal{E}=\dfrac{1}{N} \sum \limits_{i=1}^N \| \mathcal{A}_{i}-\mathcal{\hat{A}}_{i}\|_2$. Fig.~\ref{fig:figure1}(c) shows $\mathcal{E}$ as a function of $M/H$. Different values of $M/H$ have been obtained by changing the  number of measurements $M$, while the number of unknown interactions to reconstruct $H$, given the symmetry of the interaction terms in Eq.~\eqref{eq:LotKaVolterra}, is fixed to $H=N-1+(N-1)(N-2)/2 =21$.  The results indicate that our approach correctly reconstructs both pairwise and three-body interactions of the hypergraph, as the error vanishes when $M/H>1$, i.e. when the system in Eq.~\eqref{eq:linearsys} becomes overdetermined. This is evident also from Fig.~\ref{fig:figure1}(d) reporting the true values of the weights associated to non-zero interactions in the microbial ecosystem along with the values reconstructed for the case $M/H = 0.95$, when the system in Eq.~\eqref{eq:linearsys} is underdetermined, and for the case $M/H = 1.23$, when the system is overdetermined.

\section{Coupled R\"ossler oscillators}

\begin{figure*}[t]
\centering
\includegraphics[width=0.9\linewidth]{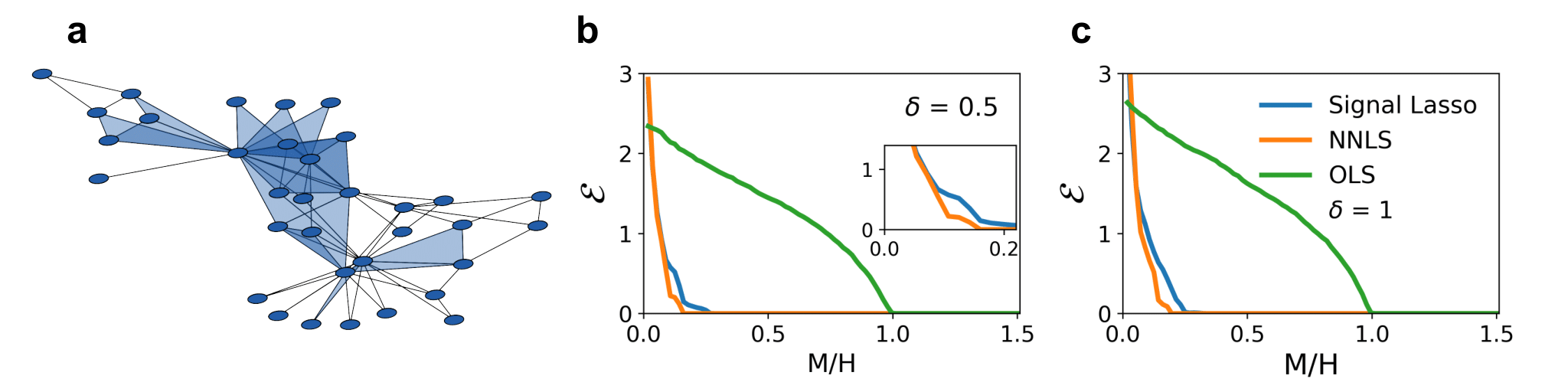}    \caption{Testing the reconstruction method on a system of $N=34$ coupled R\"ossler oscillators. (a) The underlying simplicial complex is obtained turning a tunable fraction $\delta$ of the triangles of the Zachary karate club network into 2-simplices
(case $\delta =0.5$ is shown). (b-c) Reconstruction error $\mathcal{E}$ as a function of $M/H$ for $\delta =0.5$ and $\delta =1.0$.
\label{fig:figure2}}
\end{figure*}

As a second case study we analyze the following system of R\"ossler oscillators coupled with pairwise and three-body interactions:
\begin{equation}
\label{eq:rossler}
\begin{array}{lll}
\dot{x}_{i} & = & -y_{i} -z_{i} +\sigma_1\sum\limits_{j=1}^N a_{ij}^{(1)}{g}^{(1)}(x_i,x_j)
+\sigma_2\sum\limits_{j=1}^N\sum\limits_{k=1}^N a_{ijk}^{(2)}{g}^{(2)}(x_i,x_j,x_k),\\
\dot{y}_{i} & = & x_{i} + a y_{i},  \\
\dot{z}_{i} & = & b + z_{i} (x_{i} -c),\\
\end{array}
\end{equation}
where ${g}^{(1)}(x_i,x_j)=x_j-x_i$ and ${g}^{(2)}(x_i,x_j,x_k)=x_j^2x_k-x_i^3$.
As for the underlying topology of the interactions we consider simplicial complexes constructed as follows. 
We start from the so-called Zachary karate club, which is a system originally described in terms of a graph with $N=34$ nodes and $78$ links \cite{zachary1977information}. 
Since the links form 45 triangles, we can represent the system as a simplicial complex by turning a fraction $\delta$ of the triangles into  two-dimensional simplices \cite{battiston2020networks}. By considering different values of $\delta$, we can then tune the percentage of the nodes forming a triangle which are effectively involved in a three-body interaction rather than in three, separate, pairwise interactions only.
An example of the symplicial complex obtained for $\delta=0.5$ is shown in Fig.~\ref{fig:figure2}(a). 
Given that the interactions to reconstruct are, in this case, unweighted, we can now use three different methods to solve Eq.~\eqref{eq:linearsys}. In the first method, we consider ordinary least square minimization, namely we minimize the least square error between $\mathbf{Y}_i$ and $\Phi_i \mathcal{A}_{i}$, as for the microbial ecosystem example. The green curves reported in Fig.~\ref{fig:figure2}(b) for $\delta=0.5$ and in Fig.~\ref{fig:figure2}(c) for $\delta=1$ show that the method correctly reconstructs the simplicial complex when $M/H\geq 1$.
In the second method, we consider minimization of the least square error under the additional constraint that the elements of $\mathcal{A}_i$ are non-negative: $\min_{\mathcal{A}_i \geq 0} \|\mathbf{Y}_i-\Phi_i \mathcal{A}_{i} \|_2$. 
The orange curves in Fig.~\ref{fig:figure2}(b) and (c) indicate that, including such a priori information on the nature of the interactions in the optimization problem, expands the range of values of $M/H$ for which reconstruction is possible. 
Lastly, we extend the signal lasso method \cite{shi2021inferring} to deal with higher-order interactions. Namely, we consider the following optimization: $\min ( \|\mathbf{Y}_i-\Phi_i \mathcal{A}_{i} \|_2 +\alpha \| \mathcal{A}_{i} \|_1 + \beta  \|  \mathcal{A}_{i} - \mathbf{1}_H \|_1)$, where the penalty function includes, together with the 2-norm of the difference between $\mathbf{Y}_i$ and $\Phi_i \mathcal{A}_{i}$, a regularization term to enforce sparsity of the solution, and another term to shrink the estimates to one. As indicated by the blue curves in Fig.~\ref{fig:figure2}(b) and (c), this method provides successful reconstruction in a larger range of values of $M/H$ than the ordinary method of least squares. In conclusion, by using the last two methods, we have found that a smaller sample size is sufficient to fully reconstruct the structure of the simplicial complex.  


\section{Conclusions}

Summing up, in this work we have introduced an optimization-based framework to fully reconstruct the high-order structural connectivity of a complex system. Our approach can be useful 
to understand and predict the behavior of microbial ecosystems and coupled nonlinear oscillators, and we hope that it can shed new light on a variety of physical phenomena where higher-order interactions have a fundamental role.


\begin{thebibliography}{42}%
\makeatletter
\providecommand \@ifxundefined [1]{%
 \@ifx{#1\undefined}
}%
\providecommand \@ifnum [1]{%
 \ifnum #1\expandafter \@firstoftwo
 \else \expandafter \@secondoftwo
 \fi
}%
\providecommand \@ifx [1]{%
 \ifx #1\expandafter \@firstoftwo
 \else \expandafter \@secondoftwo
 \fi
}%
\providecommand \natexlab [1]{#1}%
\providecommand \enquote  [1]{``#1''}%
\providecommand \bibnamefont  [1]{#1}%
\providecommand \bibfnamefont [1]{#1}%
\providecommand \citenamefont [1]{#1}%
\providecommand \href@noop [0]{\@secondoftwo}%
\providecommand \href [0]{\begingroup \@sanitize@url \@href}%
\providecommand \@href[1]{\@@startlink{#1}\@@href}%
\providecommand \@@href[1]{\endgroup#1\@@endlink}%
\providecommand \@sanitize@url [0]{\catcode `\\12\catcode `\$12\catcode
  `\&12\catcode `\#12\catcode `\^12\catcode `\_12\catcode `\%12\relax}%
\providecommand \@@startlink[1]{}%
\providecommand \@@endlink[0]{}%
\providecommand \url  [0]{\begingroup\@sanitize@url \@url }%
\providecommand \@url [1]{\endgroup\@href {#1}{\urlprefix }}%
\providecommand \urlprefix  [0]{URL }%
\providecommand \Eprint [0]{\href }%
\providecommand \doibase [0]{https://doi.org/}%
\providecommand \selectlanguage [0]{\@gobble}%
\providecommand \bibinfo  [0]{\@secondoftwo}%
\providecommand \bibfield  [0]{\@secondoftwo}%
\providecommand \translation [1]{[#1]}%
\providecommand \BibitemOpen [0]{}%
\providecommand \bibitemStop [0]{}%
\providecommand \bibitemNoStop [0]{.\EOS\space}%
\providecommand \EOS [0]{\spacefactor3000\relax}%
\providecommand \BibitemShut  [1]{\csname bibitem#1\endcsname}%
\let\auto@bib@innerbib\@empty
\bibitem [{\citenamefont {Grilli}\ \emph {et~al.}(2017)\citenamefont {Grilli},
  \citenamefont {Barab{\'a}s}, \citenamefont {Michalska-Smith},\ and\
  \citenamefont {Allesina}}]{grilli2017higher}%
  \BibitemOpen
  \bibfield  {author} {\bibinfo {author} {\bibfnamefont {J.}~\bibnamefont
  {Grilli}}, \bibinfo {author} {\bibfnamefont {G.}~\bibnamefont {Barab{\'a}s}},
  \bibinfo {author} {\bibfnamefont {M.~J.}\ \bibnamefont {Michalska-Smith}},\
  and\ \bibinfo {author} {\bibfnamefont {S.}~\bibnamefont {Allesina}},\
  }\bibfield  {title} {\bibinfo {title} {Higher-order interactions stabilize
  dynamics in competitive network models},\ }\href@noop {} {\bibfield
  {journal} {\bibinfo  {journal} {Nature}\ }\textbf {\bibinfo {volume} {548}},\
  \bibinfo {pages} {210} (\bibinfo {year} {2017})}\BibitemShut {NoStop}%
\bibitem [{\citenamefont {Iacopini}\ \emph {et~al.}(2019)\citenamefont
  {Iacopini}, \citenamefont {Petri}, \citenamefont {Barrat},\ and\
  \citenamefont {Latora}}]{iacopini2019simplicial}%
  \BibitemOpen
  \bibfield  {author} {\bibinfo {author} {\bibfnamefont {I.}~\bibnamefont
  {Iacopini}}, \bibinfo {author} {\bibfnamefont {G.}~\bibnamefont {Petri}},
  \bibinfo {author} {\bibfnamefont {A.}~\bibnamefont {Barrat}},\ and\ \bibinfo
  {author} {\bibfnamefont {V.}~\bibnamefont {Latora}},\ }\bibfield  {title}
  {\bibinfo {title} {Simplicial models of social contagion},\ }\href@noop {}
  {\bibfield  {journal} {\bibinfo  {journal} {Nature communications}\ }\textbf
  {\bibinfo {volume} {10}},\ \bibinfo {pages} {1} (\bibinfo {year}
  {2019})}\BibitemShut {NoStop}%
\bibitem [{\citenamefont {Yu}\ \emph {et~al.}(2011)\citenamefont {Yu},
  \citenamefont {Yang}, \citenamefont {Nakahara}, \citenamefont {Santos},
  \citenamefont {Nikoli{\'c}},\ and\ \citenamefont {Plenz}}]{yu2011higher}%
  \BibitemOpen
  \bibfield  {author} {\bibinfo {author} {\bibfnamefont {S.}~\bibnamefont
  {Yu}}, \bibinfo {author} {\bibfnamefont {H.}~\bibnamefont {Yang}}, \bibinfo
  {author} {\bibfnamefont {H.}~\bibnamefont {Nakahara}}, \bibinfo {author}
  {\bibfnamefont {G.~S.}\ \bibnamefont {Santos}}, \bibinfo {author}
  {\bibfnamefont {D.}~\bibnamefont {Nikoli{\'c}}},\ and\ \bibinfo {author}
  {\bibfnamefont {D.}~\bibnamefont {Plenz}},\ }\bibfield  {title} {\bibinfo
  {title} {Higher-order interactions characterized in cortical activity},\
  }\href@noop {} {\bibfield  {journal} {\bibinfo  {journal} {Journal of
  neuroscience}\ }\textbf {\bibinfo {volume} {31}},\ \bibinfo {pages} {17514}
  (\bibinfo {year} {2011})}\BibitemShut {NoStop}%
\bibitem [{\citenamefont {Battiston}\ \emph {et~al.}(2020)\citenamefont
  {Battiston}, \citenamefont {Cencetti}, \citenamefont {Iacopini},
  \citenamefont {Latora}, \citenamefont {Lucas}, \citenamefont {Patania},
  \citenamefont {Young},\ and\ \citenamefont {Petri}}]{battiston2020networks}%
  \BibitemOpen
  \bibfield  {author} {\bibinfo {author} {\bibfnamefont {F.}~\bibnamefont
  {Battiston}}, \bibinfo {author} {\bibfnamefont {G.}~\bibnamefont {Cencetti}},
  \bibinfo {author} {\bibfnamefont {I.}~\bibnamefont {Iacopini}}, \bibinfo
  {author} {\bibfnamefont {V.}~\bibnamefont {Latora}}, \bibinfo {author}
  {\bibfnamefont {M.}~\bibnamefont {Lucas}}, \bibinfo {author} {\bibfnamefont
  {A.}~\bibnamefont {Patania}}, \bibinfo {author} {\bibfnamefont {J.-G.}\
  \bibnamefont {Young}},\ and\ \bibinfo {author} {\bibfnamefont
  {G.}~\bibnamefont {Petri}},\ }\bibfield  {title} {\bibinfo {title} {Networks
  beyond pairwise interactions: structure and dynamics},\ }\href@noop {}
  {\bibfield  {journal} {\bibinfo  {journal} {Physics Reports}\ }\textbf
  {\bibinfo {volume} {874}},\ \bibinfo {pages} {1} (\bibinfo {year}
  {2020})}\BibitemShut {NoStop}%
\bibitem [{\citenamefont {Bianconi}(2021)}]{bianconi2021higher}%
  \BibitemOpen
  \bibfield  {author} {\bibinfo {author} {\bibfnamefont {G.}~\bibnamefont
  {Bianconi}},\ }\href@noop {} {\emph {\bibinfo {title} {Higher-order
  networks}}}\ (\bibinfo  {publisher} {Cambridge University Press},\ \bibinfo
  {year} {2021})\BibitemShut {NoStop}%
\bibitem [{\citenamefont {Hatcher}(2005)}]{hatcher2005algebraic}%
  \BibitemOpen
  \bibfield  {author} {\bibinfo {author} {\bibfnamefont {A.}~\bibnamefont
  {Hatcher}},\ }\href@noop {} {\emph {\bibinfo {title} {Algebraic topology}}}\
  (\bibinfo  {publisher} {Cambridge University Press},\ \bibinfo {year}
  {2005})\BibitemShut {NoStop}%
\bibitem [{\citenamefont {Berge}(1973)}]{berge1973graphs}%
  \BibitemOpen
  \bibfield  {author} {\bibinfo {author} {\bibfnamefont {C.}~\bibnamefont
  {Berge}},\ }\href@noop {} {\emph {\bibinfo {title} {Graphs and
  hypergraphs}}}\ (\bibinfo  {publisher} {North-Holland Pub. Co.},\ \bibinfo
  {year} {1973})\BibitemShut {NoStop}%
\bibitem [{\citenamefont {Alvarez-Rodriguez}\ \emph {et~al.}(2021)\citenamefont
  {Alvarez-Rodriguez}, \citenamefont {Battiston}, \citenamefont {de~Arruda},
  \citenamefont {Moreno}, \citenamefont {Perc},\ and\ \citenamefont
  {Latora}}]{alvarez2021evolutionary}%
  \BibitemOpen
  \bibfield  {author} {\bibinfo {author} {\bibfnamefont {U.}~\bibnamefont
  {Alvarez-Rodriguez}}, \bibinfo {author} {\bibfnamefont {F.}~\bibnamefont
  {Battiston}}, \bibinfo {author} {\bibfnamefont {G.~F.}\ \bibnamefont
  {de~Arruda}}, \bibinfo {author} {\bibfnamefont {Y.}~\bibnamefont {Moreno}},
  \bibinfo {author} {\bibfnamefont {M.}~\bibnamefont {Perc}},\ and\ \bibinfo
  {author} {\bibfnamefont {V.}~\bibnamefont {Latora}},\ }\bibfield  {title}
  {\bibinfo {title} {Evolutionary dynamics of higher-order interactions in
  social networks},\ }\href@noop {} {\bibfield  {journal} {\bibinfo  {journal}
  {Nature Human Behaviour}\ }\textbf {\bibinfo {volume} {5}},\ \bibinfo {pages}
  {586} (\bibinfo {year} {2021})}\BibitemShut {NoStop}%
\bibitem [{\citenamefont {Gambuzza}\ \emph {et~al.}(2021)\citenamefont
  {Gambuzza}, \citenamefont {Di~Patti}, \citenamefont {Gallo}, \citenamefont
  {Lepri}, \citenamefont {Romance}, \citenamefont {Criado}, \citenamefont
  {Frasca}, \citenamefont {Latora},\ and\ \citenamefont
  {Boccaletti}}]{gambuzza2021stability}%
  \BibitemOpen
  \bibfield  {author} {\bibinfo {author} {\bibfnamefont {L.~V.}\ \bibnamefont
  {Gambuzza}}, \bibinfo {author} {\bibfnamefont {F.}~\bibnamefont {Di~Patti}},
  \bibinfo {author} {\bibfnamefont {L.}~\bibnamefont {Gallo}}, \bibinfo
  {author} {\bibfnamefont {S.}~\bibnamefont {Lepri}}, \bibinfo {author}
  {\bibfnamefont {M.}~\bibnamefont {Romance}}, \bibinfo {author} {\bibfnamefont
  {R.}~\bibnamefont {Criado}}, \bibinfo {author} {\bibfnamefont
  {M.}~\bibnamefont {Frasca}}, \bibinfo {author} {\bibfnamefont
  {V.}~\bibnamefont {Latora}},\ and\ \bibinfo {author} {\bibfnamefont
  {S.}~\bibnamefont {Boccaletti}},\ }\bibfield  {title} {\bibinfo {title}
  {Stability of synchronization in simplicial complexes},\ }\href@noop {}
  {\bibfield  {journal} {\bibinfo  {journal} {Nature communications}\ }\textbf
  {\bibinfo {volume} {12}},\ \bibinfo {pages} {1} (\bibinfo {year}
  {2021})}\BibitemShut {NoStop}%
\bibitem [{\citenamefont {Gallo}\ \emph
  {et~al.}(2022{\natexlab{a}})\citenamefont {Gallo}, \citenamefont {Muolo},
  \citenamefont {Gambuzza}, \citenamefont {Latora}, \citenamefont {Frasca},\
  and\ \citenamefont {Carletti}}]{gallo2022synchronization}%
  \BibitemOpen
  \bibfield  {author} {\bibinfo {author} {\bibfnamefont {L.}~\bibnamefont
  {Gallo}}, \bibinfo {author} {\bibfnamefont {R.}~\bibnamefont {Muolo}},
  \bibinfo {author} {\bibfnamefont {L.~V.}\ \bibnamefont {Gambuzza}}, \bibinfo
  {author} {\bibfnamefont {V.}~\bibnamefont {Latora}}, \bibinfo {author}
  {\bibfnamefont {M.}~\bibnamefont {Frasca}},\ and\ \bibinfo {author}
  {\bibfnamefont {T.}~\bibnamefont {Carletti}},\ }\bibfield  {title} {\bibinfo
  {title} {Synchronization induced by directed higher-order interactions},\
  }\href@noop {} {\bibfield  {journal} {\bibinfo  {journal} {Communications
  Physics}\ }\textbf {\bibinfo {volume} {5}},\ \bibinfo {pages} {263} (\bibinfo
  {year} {2022}{\natexlab{a}})}\BibitemShut {NoStop}%
\bibitem [{\citenamefont {Muolo}\ \emph {et~al.}(2023)\citenamefont {Muolo},
  \citenamefont {Gallo}, \citenamefont {Latora}, \citenamefont {Frasca},\ and\
  \citenamefont {Carletti}}]{muolo2023turing}%
  \BibitemOpen
  \bibfield  {author} {\bibinfo {author} {\bibfnamefont {R.}~\bibnamefont
  {Muolo}}, \bibinfo {author} {\bibfnamefont {L.}~\bibnamefont {Gallo}},
  \bibinfo {author} {\bibfnamefont {V.}~\bibnamefont {Latora}}, \bibinfo
  {author} {\bibfnamefont {M.}~\bibnamefont {Frasca}},\ and\ \bibinfo {author}
  {\bibfnamefont {T.}~\bibnamefont {Carletti}},\ }\bibfield  {title} {\bibinfo
  {title} {Turing patterns in systems with high-order interactions},\
  }\href@noop {} {\bibfield  {journal} {\bibinfo  {journal} {Chaos, Solitons \&
  Fractals}\ }\textbf {\bibinfo {volume} {166}},\ \bibinfo {pages} {112912}
  (\bibinfo {year} {2023})}\BibitemShut {NoStop}%
\bibitem [{\citenamefont {Battiston}\ \emph {et~al.}(2021)\citenamefont
  {Battiston}, \citenamefont {Amico}, \citenamefont {Barrat}, \citenamefont
  {Bianconi}, \citenamefont {Ferraz~de Arruda}, \citenamefont {Franceschiello},
  \citenamefont {Iacopini}, \citenamefont {K{\'e}fi}, \citenamefont {Latora},
  \citenamefont {Moreno} \emph {et~al.}}]{battiston2021physics}%
  \BibitemOpen
  \bibfield  {author} {\bibinfo {author} {\bibfnamefont {F.}~\bibnamefont
  {Battiston}}, \bibinfo {author} {\bibfnamefont {E.}~\bibnamefont {Amico}},
  \bibinfo {author} {\bibfnamefont {A.}~\bibnamefont {Barrat}}, \bibinfo
  {author} {\bibfnamefont {G.}~\bibnamefont {Bianconi}}, \bibinfo {author}
  {\bibfnamefont {G.}~\bibnamefont {Ferraz~de Arruda}}, \bibinfo {author}
  {\bibfnamefont {B.}~\bibnamefont {Franceschiello}}, \bibinfo {author}
  {\bibfnamefont {I.}~\bibnamefont {Iacopini}}, \bibinfo {author}
  {\bibfnamefont {S.}~\bibnamefont {K{\'e}fi}}, \bibinfo {author}
  {\bibfnamefont {V.}~\bibnamefont {Latora}}, \bibinfo {author} {\bibfnamefont
  {Y.}~\bibnamefont {Moreno}}, \emph {et~al.},\ }\bibfield  {title} {\bibinfo
  {title} {The physics of higher-order interactions in complex systems},\
  }\href@noop {} {\bibfield  {journal} {\bibinfo  {journal} {Nature Physics}\
  }\textbf {\bibinfo {volume} {17}},\ \bibinfo {pages} {1093} (\bibinfo {year}
  {2021})}\BibitemShut {NoStop}%
\bibitem [{\citenamefont {Battiston}\ and\ \citenamefont
  {Petri}(2022)}]{battiston2022higher}%
  \BibitemOpen
  \bibfield  {author} {\bibinfo {author} {\bibfnamefont {F.}~\bibnamefont
  {Battiston}}\ and\ \bibinfo {author} {\bibfnamefont {G.}~\bibnamefont
  {Petri}},\ }\href@noop {} {\emph {\bibinfo {title} {Higher-Order Systems}}}\
  (\bibinfo  {publisher} {Springer},\ \bibinfo {year} {2022})\BibitemShut
  {NoStop}%
\bibitem [{\citenamefont {Timme}\ and\ \citenamefont
  {Casadiego}(2014)}]{timme2014revealing}%
  \BibitemOpen
  \bibfield  {author} {\bibinfo {author} {\bibfnamefont {M.}~\bibnamefont
  {Timme}}\ and\ \bibinfo {author} {\bibfnamefont {J.}~\bibnamefont
  {Casadiego}},\ }\bibfield  {title} {\bibinfo {title} {Revealing networks from
  dynamics: an introduction},\ }\href@noop {} {\bibfield  {journal} {\bibinfo
  {journal} {Journal of Physics A: Mathematical and Theoretical}\ }\textbf
  {\bibinfo {volume} {47}},\ \bibinfo {pages} {343001} (\bibinfo {year}
  {2014})}\BibitemShut {NoStop}%
\bibitem [{\citenamefont {Ren}\ \emph {et~al.}(2010)\citenamefont {Ren},
  \citenamefont {Wang}, \citenamefont {Li},\ and\ \citenamefont
  {Lai}}]{ren2010noise}%
  \BibitemOpen
  \bibfield  {author} {\bibinfo {author} {\bibfnamefont {J.}~\bibnamefont
  {Ren}}, \bibinfo {author} {\bibfnamefont {W.-X.}\ \bibnamefont {Wang}},
  \bibinfo {author} {\bibfnamefont {B.}~\bibnamefont {Li}},\ and\ \bibinfo
  {author} {\bibfnamefont {Y.-C.}\ \bibnamefont {Lai}},\ }\bibfield  {title}
  {\bibinfo {title} {Noise bridges dynamical correlation and topology in
  coupled oscillator networks},\ }\href@noop {} {\bibfield  {journal} {\bibinfo
   {journal} {Physical review letters}\ }\textbf {\bibinfo {volume} {104}},\
  \bibinfo {pages} {058701} (\bibinfo {year} {2010})}\BibitemShut {NoStop}%
\bibitem [{\citenamefont {Wu}\ \emph {et~al.}(2011)\citenamefont {Wu},
  \citenamefont {Zhou}, \citenamefont {Chen},\ and\ \citenamefont
  {Lu}}]{wu2011detecting}%
  \BibitemOpen
  \bibfield  {author} {\bibinfo {author} {\bibfnamefont {X.}~\bibnamefont
  {Wu}}, \bibinfo {author} {\bibfnamefont {C.}~\bibnamefont {Zhou}}, \bibinfo
  {author} {\bibfnamefont {G.}~\bibnamefont {Chen}},\ and\ \bibinfo {author}
  {\bibfnamefont {J.-a.}\ \bibnamefont {Lu}},\ }\bibfield  {title} {\bibinfo
  {title} {Detecting the topologies of complex networks with stochastic
  perturbations},\ }\href@noop {} {\bibfield  {journal} {\bibinfo  {journal}
  {Chaos: An Interdisciplinary Journal of Nonlinear Science}\ }\textbf
  {\bibinfo {volume} {21}},\ \bibinfo {pages} {043129} (\bibinfo {year}
  {2011})}\BibitemShut {NoStop}%
\bibitem [{\citenamefont {Jansen}\ \emph {et~al.}(2003)\citenamefont {Jansen},
  \citenamefont {Yu}, \citenamefont {Greenbaum}, \citenamefont {Kluger},
  \citenamefont {Krogan}, \citenamefont {Chung}, \citenamefont {Emili},
  \citenamefont {Snyder}, \citenamefont {Greenblatt},\ and\ \citenamefont
  {Gerstein}}]{jansen2003bayesian}%
  \BibitemOpen
  \bibfield  {author} {\bibinfo {author} {\bibfnamefont {R.}~\bibnamefont
  {Jansen}}, \bibinfo {author} {\bibfnamefont {H.}~\bibnamefont {Yu}}, \bibinfo
  {author} {\bibfnamefont {D.}~\bibnamefont {Greenbaum}}, \bibinfo {author}
  {\bibfnamefont {Y.}~\bibnamefont {Kluger}}, \bibinfo {author} {\bibfnamefont
  {N.~J.}\ \bibnamefont {Krogan}}, \bibinfo {author} {\bibfnamefont
  {S.}~\bibnamefont {Chung}}, \bibinfo {author} {\bibfnamefont
  {A.}~\bibnamefont {Emili}}, \bibinfo {author} {\bibfnamefont
  {M.}~\bibnamefont {Snyder}}, \bibinfo {author} {\bibfnamefont {J.~F.}\
  \bibnamefont {Greenblatt}},\ and\ \bibinfo {author} {\bibfnamefont
  {M.}~\bibnamefont {Gerstein}},\ }\bibfield  {title} {\bibinfo {title} {A
  bayesian networks approach for predicting protein-protein interactions from
  genomic data},\ }\href@noop {} {\bibfield  {journal} {\bibinfo  {journal}
  {science}\ }\textbf {\bibinfo {volume} {302}},\ \bibinfo {pages} {449}
  (\bibinfo {year} {2003})}\BibitemShut {NoStop}%
\bibitem [{\citenamefont {Timme}(2007)}]{timme2007revealing}%
  \BibitemOpen
  \bibfield  {author} {\bibinfo {author} {\bibfnamefont {M.}~\bibnamefont
  {Timme}},\ }\bibfield  {title} {\bibinfo {title} {Revealing network
  connectivity from response dynamics},\ }\href@noop {} {\bibfield  {journal}
  {\bibinfo  {journal} {Physical Review Letters}\ }\textbf {\bibinfo {volume}
  {98}},\ \bibinfo {pages} {224101} (\bibinfo {year} {2007})}\BibitemShut
  {NoStop}%
\bibitem [{\citenamefont {Yu}\ \emph {et~al.}(2006)\citenamefont {Yu},
  \citenamefont {Righero},\ and\ \citenamefont {Kocarev}}]{yu2006estimating}%
  \BibitemOpen
  \bibfield  {author} {\bibinfo {author} {\bibfnamefont {D.}~\bibnamefont
  {Yu}}, \bibinfo {author} {\bibfnamefont {M.}~\bibnamefont {Righero}},\ and\
  \bibinfo {author} {\bibfnamefont {L.}~\bibnamefont {Kocarev}},\ }\bibfield
  {title} {\bibinfo {title} {Estimating topology of networks},\ }\href@noop {}
  {\bibfield  {journal} {\bibinfo  {journal} {Physical Review Letters}\
  }\textbf {\bibinfo {volume} {97}},\ \bibinfo {pages} {188701} (\bibinfo
  {year} {2006})}\BibitemShut {NoStop}%
\bibitem [{\citenamefont {Wu}\ \emph {et~al.}(2015)\citenamefont {Wu},
  \citenamefont {Zhao}, \citenamefont {L{\"u}}, \citenamefont {Tang},\ and\
  \citenamefont {Lu}}]{wu2015identifying}%
  \BibitemOpen
  \bibfield  {author} {\bibinfo {author} {\bibfnamefont {X.}~\bibnamefont
  {Wu}}, \bibinfo {author} {\bibfnamefont {X.}~\bibnamefont {Zhao}}, \bibinfo
  {author} {\bibfnamefont {J.}~\bibnamefont {L{\"u}}}, \bibinfo {author}
  {\bibfnamefont {L.}~\bibnamefont {Tang}},\ and\ \bibinfo {author}
  {\bibfnamefont {J.-a.}\ \bibnamefont {Lu}},\ }\bibfield  {title} {\bibinfo
  {title} {Identifying topologies of complex dynamical networks with stochastic
  perturbations},\ }\href@noop {} {\bibfield  {journal} {\bibinfo  {journal}
  {IEEE Transactions on Control of Network Systems}\ }\textbf {\bibinfo
  {volume} {3}},\ \bibinfo {pages} {379} (\bibinfo {year} {2015})}\BibitemShut
  {NoStop}%
\bibitem [{\citenamefont {Shandilya}\ and\ \citenamefont
  {Timme}(2011)}]{shandilya2011inferring}%
  \BibitemOpen
  \bibfield  {author} {\bibinfo {author} {\bibfnamefont {S.~G.}\ \bibnamefont
  {Shandilya}}\ and\ \bibinfo {author} {\bibfnamefont {M.}~\bibnamefont
  {Timme}},\ }\bibfield  {title} {\bibinfo {title} {Inferring network topology
  from complex dynamics},\ }\href@noop {} {\bibfield  {journal} {\bibinfo
  {journal} {New Journal of Physics}\ }\textbf {\bibinfo {volume} {13}},\
  \bibinfo {pages} {013004} (\bibinfo {year} {2011})}\BibitemShut {NoStop}%
\bibitem [{\citenamefont {Han}\ \emph {et~al.}(2015)\citenamefont {Han},
  \citenamefont {Shen}, \citenamefont {Wang},\ and\ \citenamefont
  {Di}}]{han2015robust}%
  \BibitemOpen
  \bibfield  {author} {\bibinfo {author} {\bibfnamefont {X.}~\bibnamefont
  {Han}}, \bibinfo {author} {\bibfnamefont {Z.}~\bibnamefont {Shen}}, \bibinfo
  {author} {\bibfnamefont {W.-X.}\ \bibnamefont {Wang}},\ and\ \bibinfo
  {author} {\bibfnamefont {Z.}~\bibnamefont {Di}},\ }\bibfield  {title}
  {\bibinfo {title} {Robust reconstruction of complex networks from sparse
  data},\ }\href@noop {} {\bibfield  {journal} {\bibinfo  {journal} {Physical
  review letters}\ }\textbf {\bibinfo {volume} {114}},\ \bibinfo {pages}
  {028701} (\bibinfo {year} {2015})}\BibitemShut {NoStop}%
\bibitem [{\citenamefont {Shi}\ \emph {et~al.}(2021)\citenamefont {Shi},
  \citenamefont {Shen}, \citenamefont {Jin}, \citenamefont {Shi}, \citenamefont
  {Wang},\ and\ \citenamefont {Boccaletti}}]{shi2021inferring}%
  \BibitemOpen
  \bibfield  {author} {\bibinfo {author} {\bibfnamefont {L.}~\bibnamefont
  {Shi}}, \bibinfo {author} {\bibfnamefont {C.}~\bibnamefont {Shen}}, \bibinfo
  {author} {\bibfnamefont {L.}~\bibnamefont {Jin}}, \bibinfo {author}
  {\bibfnamefont {Q.}~\bibnamefont {Shi}}, \bibinfo {author} {\bibfnamefont
  {Z.}~\bibnamefont {Wang}},\ and\ \bibinfo {author} {\bibfnamefont
  {S.}~\bibnamefont {Boccaletti}},\ }\bibfield  {title} {\bibinfo {title}
  {Inferring network structures via signal lasso},\ }\href@noop {} {\bibfield
  {journal} {\bibinfo  {journal} {Physical Review Research}\ }\textbf {\bibinfo
  {volume} {3}},\ \bibinfo {pages} {043210} (\bibinfo {year}
  {2021})}\BibitemShut {NoStop}%
\bibitem [{\citenamefont {Rosas}\ \emph {et~al.}(2022)\citenamefont {Rosas},
  \citenamefont {Mediano}, \citenamefont {Luppi}, \citenamefont {Varley},
  \citenamefont {Lizier}, \citenamefont {Stramaglia}, \citenamefont {Jensen},\
  and\ \citenamefont {Marinazzo}}]{rosas2022disentangling}%
  \BibitemOpen
  \bibfield  {author} {\bibinfo {author} {\bibfnamefont {F.~E.}\ \bibnamefont
  {Rosas}}, \bibinfo {author} {\bibfnamefont {P.~A.}\ \bibnamefont {Mediano}},
  \bibinfo {author} {\bibfnamefont {A.~I.}\ \bibnamefont {Luppi}}, \bibinfo
  {author} {\bibfnamefont {T.~F.}\ \bibnamefont {Varley}}, \bibinfo {author}
  {\bibfnamefont {J.~T.}\ \bibnamefont {Lizier}}, \bibinfo {author}
  {\bibfnamefont {S.}~\bibnamefont {Stramaglia}}, \bibinfo {author}
  {\bibfnamefont {H.~J.}\ \bibnamefont {Jensen}},\ and\ \bibinfo {author}
  {\bibfnamefont {D.}~\bibnamefont {Marinazzo}},\ }\bibfield  {title} {\bibinfo
  {title} {Disentangling high-order mechanisms and high-order behaviours in
  complex systems},\ }\href@noop {} {\bibfield  {journal} {\bibinfo  {journal}
  {Nature Physics}\ }\textbf {\bibinfo {volume} {18}},\ \bibinfo {pages} {476}
  (\bibinfo {year} {2022})}\BibitemShut {NoStop}%
\bibitem [{\citenamefont {Marinazzo}\ \emph {et~al.}(2022)\citenamefont
  {Marinazzo}, \citenamefont {Van~Roozendaal}, \citenamefont {Rosas},
  \citenamefont {Stella}, \citenamefont {Comolatti}, \citenamefont {Colenbier},
  \citenamefont {Stramaglia},\ and\ \citenamefont
  {Rosseel}}]{marinazzo2022information}%
  \BibitemOpen
  \bibfield  {author} {\bibinfo {author} {\bibfnamefont {D.}~\bibnamefont
  {Marinazzo}}, \bibinfo {author} {\bibfnamefont {J.}~\bibnamefont
  {Van~Roozendaal}}, \bibinfo {author} {\bibfnamefont {F.~E.}\ \bibnamefont
  {Rosas}}, \bibinfo {author} {\bibfnamefont {M.}~\bibnamefont {Stella}},
  \bibinfo {author} {\bibfnamefont {R.}~\bibnamefont {Comolatti}}, \bibinfo
  {author} {\bibfnamefont {N.}~\bibnamefont {Colenbier}}, \bibinfo {author}
  {\bibfnamefont {S.}~\bibnamefont {Stramaglia}},\ and\ \bibinfo {author}
  {\bibfnamefont {Y.}~\bibnamefont {Rosseel}},\ }\bibfield  {title} {\bibinfo
  {title} {An information-theoretic approach to hypergraph psychometrics},\
  }\href@noop {} {\bibfield  {journal} {\bibinfo  {journal} {arXiv preprint
  arXiv:2205.01035}\ } (\bibinfo {year} {2022})}\BibitemShut {NoStop}%
\bibitem [{\citenamefont {Pernice}\ \emph {et~al.}(2022)\citenamefont
  {Pernice}, \citenamefont {Faes}, \citenamefont {Feucht}, \citenamefont
  {Benninger}, \citenamefont {Mangione},\ and\ \citenamefont
  {Schiecke}}]{pernice2022pairwise}%
  \BibitemOpen
  \bibfield  {author} {\bibinfo {author} {\bibfnamefont {R.}~\bibnamefont
  {Pernice}}, \bibinfo {author} {\bibfnamefont {L.}~\bibnamefont {Faes}},
  \bibinfo {author} {\bibfnamefont {M.}~\bibnamefont {Feucht}}, \bibinfo
  {author} {\bibfnamefont {F.}~\bibnamefont {Benninger}}, \bibinfo {author}
  {\bibfnamefont {S.}~\bibnamefont {Mangione}},\ and\ \bibinfo {author}
  {\bibfnamefont {K.}~\bibnamefont {Schiecke}},\ }\bibfield  {title} {\bibinfo
  {title} {Pairwise and higher-order measures of brain-heart interactions in
  children with temporal lobe epilepsy},\ }\href@noop {} {\bibfield  {journal}
  {\bibinfo  {journal} {Journal of Neural Engineering}\ } (\bibinfo {year}
  {2022})}\BibitemShut {NoStop}%
\bibitem [{\citenamefont {Santoro}\ \emph {et~al.}(2023)\citenamefont
  {Santoro}, \citenamefont {Battiston}, \citenamefont {Petri},\ and\
  \citenamefont {Amico}}]{santoro2023unveiling}%
  \BibitemOpen
  \bibfield  {author} {\bibinfo {author} {\bibfnamefont {A.}~\bibnamefont
  {Santoro}}, \bibinfo {author} {\bibfnamefont {F.}~\bibnamefont {Battiston}},
  \bibinfo {author} {\bibfnamefont {G.}~\bibnamefont {Petri}},\ and\ \bibinfo
  {author} {\bibfnamefont {E.}~\bibnamefont {Amico}},\ }\bibfield  {title}
  {\bibinfo {title} {Higher-order organization of multivariate time series},\
  }\href@noop {} {\bibfield  {journal} {\bibinfo  {journal} {Nature Physics}\ }
  (\bibinfo {year} {2023})}\BibitemShut {NoStop}%
\bibitem [{\citenamefont {Musciotto}\ \emph {et~al.}(2021)\citenamefont
  {Musciotto}, \citenamefont {Battiston},\ and\ \citenamefont
  {Mantegna}}]{musciotto2021detecting}%
  \BibitemOpen
  \bibfield  {author} {\bibinfo {author} {\bibfnamefont {F.}~\bibnamefont
  {Musciotto}}, \bibinfo {author} {\bibfnamefont {F.}~\bibnamefont
  {Battiston}},\ and\ \bibinfo {author} {\bibfnamefont {R.~N.}\ \bibnamefont
  {Mantegna}},\ }\bibfield  {title} {\bibinfo {title} {Detecting informative
  higher-order interactions in statistically validated hypergraphs},\
  }\href@noop {} {\bibfield  {journal} {\bibinfo  {journal} {Communications
  Physics}\ }\textbf {\bibinfo {volume} {4}},\ \bibinfo {pages} {1} (\bibinfo
  {year} {2021})}\BibitemShut {NoStop}%
\bibitem [{\citenamefont {Musciotto}\ \emph {et~al.}(2022)\citenamefont
  {Musciotto}, \citenamefont {Battiston},\ and\ \citenamefont
  {Mantegna}}]{musciotto2022identifying}%
  \BibitemOpen
  \bibfield  {author} {\bibinfo {author} {\bibfnamefont {F.}~\bibnamefont
  {Musciotto}}, \bibinfo {author} {\bibfnamefont {F.}~\bibnamefont
  {Battiston}},\ and\ \bibinfo {author} {\bibfnamefont {R.~N.}\ \bibnamefont
  {Mantegna}},\ }\bibfield  {title} {\bibinfo {title} {Identifying maximal sets
  of significantly interacting nodes in higher-order networks},\ }\href@noop {}
  {\bibfield  {journal} {\bibinfo  {journal} {arXiv preprint arXiv:2209.12712}\
  } (\bibinfo {year} {2022})}\BibitemShut {NoStop}%
\bibitem [{\citenamefont {Young}\ \emph {et~al.}(2021)\citenamefont {Young},
  \citenamefont {Petri},\ and\ \citenamefont {Peixoto}}]{young2021hypergraph}%
  \BibitemOpen
  \bibfield  {author} {\bibinfo {author} {\bibfnamefont {J.-G.}\ \bibnamefont
  {Young}}, \bibinfo {author} {\bibfnamefont {G.}~\bibnamefont {Petri}},\ and\
  \bibinfo {author} {\bibfnamefont {T.~P.}\ \bibnamefont {Peixoto}},\
  }\bibfield  {title} {\bibinfo {title} {Hypergraph reconstruction from network
  data},\ }\href@noop {} {\bibfield  {journal} {\bibinfo  {journal}
  {Communications Physics}\ }\textbf {\bibinfo {volume} {4}},\ \bibinfo {pages}
  {1} (\bibinfo {year} {2021})}\BibitemShut {NoStop}%
\bibitem [{\citenamefont {Lizotte}\ \emph {et~al.}(2022)\citenamefont
  {Lizotte}, \citenamefont {Young},\ and\ \citenamefont
  {Allard}}]{lizotte2022hypergraph}%
  \BibitemOpen
  \bibfield  {author} {\bibinfo {author} {\bibfnamefont {S.}~\bibnamefont
  {Lizotte}}, \bibinfo {author} {\bibfnamefont {J.-G.}\ \bibnamefont {Young}},\
  and\ \bibinfo {author} {\bibfnamefont {A.}~\bibnamefont {Allard}},\
  }\bibfield  {title} {\bibinfo {title} {Hypergraph reconstruction from noisy
  pairwise observations},\ }\href@noop {} {\bibfield  {journal} {\bibinfo
  {journal} {arXiv preprint arXiv:2208.06503}\ } (\bibinfo {year}
  {2022})}\BibitemShut {NoStop}%
\bibitem [{\citenamefont {Wang}\ \emph {et~al.}(2022)\citenamefont {Wang},
  \citenamefont {Ma}, \citenamefont {Chen}, \citenamefont {Lai},\ and\
  \citenamefont {Zhang}}]{wang2022full}%
  \BibitemOpen
  \bibfield  {author} {\bibinfo {author} {\bibfnamefont {H.}~\bibnamefont
  {Wang}}, \bibinfo {author} {\bibfnamefont {C.}~\bibnamefont {Ma}}, \bibinfo
  {author} {\bibfnamefont {H.-S.}\ \bibnamefont {Chen}}, \bibinfo {author}
  {\bibfnamefont {Y.-C.}\ \bibnamefont {Lai}},\ and\ \bibinfo {author}
  {\bibfnamefont {H.-F.}\ \bibnamefont {Zhang}},\ }\bibfield  {title} {\bibinfo
  {title} {Full reconstruction of simplicial complexes from binary contagion
  and {I}sing data},\ }\href@noop {} {\bibfield  {journal} {\bibinfo  {journal}
  {Nature Communications}\ }\textbf {\bibinfo {volume} {13}},\ \bibinfo {pages}
  {1} (\bibinfo {year} {2022})}\BibitemShut {NoStop}%
\bibitem [{\citenamefont {Prasse}\ and\ \citenamefont
  {Van~Mieghem}(2022)}]{prasse2022predicting}%
  \BibitemOpen
  \bibfield  {author} {\bibinfo {author} {\bibfnamefont {B.}~\bibnamefont
  {Prasse}}\ and\ \bibinfo {author} {\bibfnamefont {P.}~\bibnamefont
  {Van~Mieghem}},\ }\bibfield  {title} {\bibinfo {title} {Predicting network
  dynamics without requiring the knowledge of the interaction graph},\
  }\href@noop {} {\bibfield  {journal} {\bibinfo  {journal} {Proceedings of the
  National Academy of Sciences}\ }\textbf {\bibinfo {volume} {119}},\ \bibinfo
  {pages} {e2205517119} (\bibinfo {year} {2022})}\BibitemShut {NoStop}%
\bibitem [{\citenamefont {Keesman}\ and\ \citenamefont
  {Keesman}(2011)}]{keesman2011system}%
  \BibitemOpen
  \bibfield  {author} {\bibinfo {author} {\bibfnamefont {K.~J.}\ \bibnamefont
  {Keesman}}\ and\ \bibinfo {author} {\bibfnamefont {K.~J.}\ \bibnamefont
  {Keesman}},\ }\href@noop {} {\emph {\bibinfo {title} {System identification:
  an introduction}}},\ Vol.~\bibinfo {volume} {2}\ (\bibinfo  {publisher}
  {Springer},\ \bibinfo {year} {2011})\BibitemShut {NoStop}%
\bibitem [{\citenamefont {Gallo}\ \emph
  {et~al.}(2022{\natexlab{b}})\citenamefont {Gallo}, \citenamefont {Frasca},
  \citenamefont {Latora},\ and\ \citenamefont {Russo}}]{gallo2022lack}%
  \BibitemOpen
  \bibfield  {author} {\bibinfo {author} {\bibfnamefont {L.}~\bibnamefont
  {Gallo}}, \bibinfo {author} {\bibfnamefont {M.}~\bibnamefont {Frasca}},
  \bibinfo {author} {\bibfnamefont {V.}~\bibnamefont {Latora}},\ and\ \bibinfo
  {author} {\bibfnamefont {G.}~\bibnamefont {Russo}},\ }\bibfield  {title}
  {\bibinfo {title} {Lack of practical identifiability may hamper reliable
  predictions in covid-19 epidemic models},\ }\href@noop {} {\bibfield
  {journal} {\bibinfo  {journal} {Science advances}\ }\textbf {\bibinfo
  {volume} {8}},\ \bibinfo {pages} {eabg5234} (\bibinfo {year}
  {2022}{\natexlab{b}})}\BibitemShut {NoStop}%
\bibitem [{\citenamefont {Hibbing}\ \emph {et~al.}(2010)\citenamefont
  {Hibbing}, \citenamefont {Fuqua}, \citenamefont {Parsek},\ and\ \citenamefont
  {Peterson}}]{hibbing2010bacterial}%
  \BibitemOpen
  \bibfield  {author} {\bibinfo {author} {\bibfnamefont {M.~E.}\ \bibnamefont
  {Hibbing}}, \bibinfo {author} {\bibfnamefont {C.}~\bibnamefont {Fuqua}},
  \bibinfo {author} {\bibfnamefont {M.~R.}\ \bibnamefont {Parsek}},\ and\
  \bibinfo {author} {\bibfnamefont {S.~B.}\ \bibnamefont {Peterson}},\
  }\bibfield  {title} {\bibinfo {title} {Bacterial competition: surviving and
  thriving in the microbial jungle},\ }\href@noop {} {\bibfield  {journal}
  {\bibinfo  {journal} {Nature Reviews Microbiology}\ }\textbf {\bibinfo
  {volume} {8}},\ \bibinfo {pages} {15} (\bibinfo {year} {2010})}\BibitemShut
  {NoStop}%
\bibitem [{\citenamefont {Trosvik}\ \emph {et~al.}(2010)\citenamefont
  {Trosvik}, \citenamefont {Rudi}, \citenamefont {Str{\ae}tkvern},
  \citenamefont {Jakobsen}, \citenamefont {N{\ae}s},\ and\ \citenamefont
  {Stenseth}}]{trosvik2010web}%
  \BibitemOpen
  \bibfield  {author} {\bibinfo {author} {\bibfnamefont {P.}~\bibnamefont
  {Trosvik}}, \bibinfo {author} {\bibfnamefont {K.}~\bibnamefont {Rudi}},
  \bibinfo {author} {\bibfnamefont {K.~O.}\ \bibnamefont {Str{\ae}tkvern}},
  \bibinfo {author} {\bibfnamefont {K.~S.}\ \bibnamefont {Jakobsen}}, \bibinfo
  {author} {\bibfnamefont {T.}~\bibnamefont {N{\ae}s}},\ and\ \bibinfo {author}
  {\bibfnamefont {N.~C.}\ \bibnamefont {Stenseth}},\ }\bibfield  {title}
  {\bibinfo {title} {Web of ecological interactions in an experimental gut
  microbiota},\ }\href@noop {} {\bibfield  {journal} {\bibinfo  {journal}
  {Environmental microbiology}\ }\textbf {\bibinfo {volume} {12}},\ \bibinfo
  {pages} {2677} (\bibinfo {year} {2010})}\BibitemShut {NoStop}%
\bibitem [{\citenamefont {Berry}\ and\ \citenamefont
  {Widder}(2014)}]{berry2014deciphering}%
  \BibitemOpen
  \bibfield  {author} {\bibinfo {author} {\bibfnamefont {D.}~\bibnamefont
  {Berry}}\ and\ \bibinfo {author} {\bibfnamefont {S.}~\bibnamefont {Widder}},\
  }\bibfield  {title} {\bibinfo {title} {Deciphering microbial interactions and
  detecting keystone species with co-occurrence networks},\ }\href@noop {}
  {\bibfield  {journal} {\bibinfo  {journal} {Frontiers in microbiology}\
  }\textbf {\bibinfo {volume} {5}},\ \bibinfo {pages} {219} (\bibinfo {year}
  {2014})}\BibitemShut {NoStop}%
\bibitem [{\citenamefont {AlAdwani}\ and\ \citenamefont
  {Saavedra}(2019)}]{aladwani2019addition}%
  \BibitemOpen
  \bibfield  {author} {\bibinfo {author} {\bibfnamefont {M.}~\bibnamefont
  {AlAdwani}}\ and\ \bibinfo {author} {\bibfnamefont {S.}~\bibnamefont
  {Saavedra}},\ }\bibfield  {title} {\bibinfo {title} {Is the addition of
  higher-order interactions in ecological models increasing the understanding
  of ecological dynamics?},\ }\href@noop {} {\bibfield  {journal} {\bibinfo
  {journal} {Mathematical Biosciences}\ }\textbf {\bibinfo {volume} {315}},\
  \bibinfo {pages} {108222} (\bibinfo {year} {2019})}\BibitemShut {NoStop}%
\bibitem [{\citenamefont {Singh}\ and\ \citenamefont
  {Baruah}(2021)}]{singh2021higher}%
  \BibitemOpen
  \bibfield  {author} {\bibinfo {author} {\bibfnamefont {P.}~\bibnamefont
  {Singh}}\ and\ \bibinfo {author} {\bibfnamefont {G.}~\bibnamefont {Baruah}},\
  }\bibfield  {title} {\bibinfo {title} {Higher order interactions and species
  coexistence},\ }\href@noop {} {\bibfield  {journal} {\bibinfo  {journal}
  {Theoretical Ecology}\ }\textbf {\bibinfo {volume} {14}},\ \bibinfo {pages}
  {71} (\bibinfo {year} {2021})}\BibitemShut {NoStop}%
\bibitem [{\citenamefont {Case}(1999)}]{case1999illustrated}%
  \BibitemOpen
  \bibfield  {author} {\bibinfo {author} {\bibfnamefont {T.~J.}\ \bibnamefont
  {Case}},\ }\bibfield  {title} {\bibinfo {title} {Illustrated guide to
  theoretical ecology},\ }\href@noop {} {\bibfield  {journal} {\bibinfo
  {journal} {Ecology}\ }\textbf {\bibinfo {volume} {80}},\ \bibinfo {pages}
  {2848} (\bibinfo {year} {1999})}\BibitemShut {NoStop}%
\bibitem [{\citenamefont {Zachary}(1977)}]{zachary1977information}%
  \BibitemOpen
  \bibfield  {author} {\bibinfo {author} {\bibfnamefont {W.~W.}\ \bibnamefont
  {Zachary}},\ }\bibfield  {title} {\bibinfo {title} {An information flow model
  for conflict and fission in small groups},\ }\href@noop {} {\bibfield
  {journal} {\bibinfo  {journal} {Journal of anthropological research}\
  }\textbf {\bibinfo {volume} {33}},\ \bibinfo {pages} {452} (\bibinfo {year}
  {1977})}\BibitemShut {NoStop}%
\end{thebibliography}
\end{document}